\documentclass[12pt]{article}
\usepackage{epsfig}
\usepackage{axodraw}
\renewcommand{\title}{        Higher Dimensional Operators in Top Condensation
\\                              from a Renormalization Group Point of View     
}

\advance\textheight by \topskip
\renewcommand{\baselinestretch}{1.2}
\renewcommand{\thefootnote}{\fnsymbol{footnote}}

\newcommand{\beq}{\begin{equation}}
\newcommand{\eeq}{\end{equation}}
\newcommand{\bea}{\begin{eqnarray}}
\newcommand{\eea}{\end{eqnarray}}
\def\slash#1{#1\!\!\!/\!\,\,}

\def\SM{standard model }

\newcommand{\bild}[3]{
  \begin{minipage}[t]{7.5cm}
  \begin{center}\begin{picture}(200,200)
  \epsfig{file=#1.ps,height=4cm,width=4cm,
          bbllx=3.2cm ,bblly=3.2cm ,bburx=11.5cm ,bbury=11.5cm ,angle=0}
  \end{picture}\end{center}
  \vspace*{-8cm}
  \begin{center}\begin{picture}(200,200)
    #2
  \end{picture}\end{center}
  \begin{fig} #3 \label{#1} \end{fig}
  \end{minipage}}

\begin{document}
\newtheorem{fig}[figure]{Figure}


\renewcommand{\baselinestretch}{1}
\renewcommand{\thefootnote}{\alph{footnote}}

\thispagestyle{empty}

\vspace*{-0.3cm} {\bf \hfill LMU--12/97 } 

\vspace*{-0.3cm} {\bf \hfill UWTHPH--1997-36 }                 

\vspace*{-0.3cm} {\bf \hfill TUM--HEP--294/97 }  
                 
\vspace*{-0.3cm} {\bf \hfill October 97} \vspace*{1.5cm}
{\Large\bf \begin{center} \title \end{center}}
{\begin{center}

\vspace*{0.5cm}
   {\begin{center} {\large\sc
                Andreas Blumhofer\footnote{\makebox[1.cm]{Email:}
                                  Blumhofer@Photon.HEP.Physik.Uni-Muenchen.DE},
                Richard Dawid\footnote{\makebox[1.cm]{Email:}
                                  Richard.Dawid@Merlin.PAP.Univie.AC.AT}
                and Johannes Manus\footnote{\makebox[1.cm]{Email:}
                                  Johannes.Manus@Physik.TU-Muenchen.DE}}
    \end{center} }
\vspace*{0cm} {\it \begin{center}
    \footnotemark[1]Sektion Physik, Ludwig--Maximilians--Universit\"at
    M\"unchen, \\ Theresienstr.37, D--80333 M\"unchen, Germany
\\[0.5cm]
    \footnotemark[2]Institut f\"ur Theoretische Physik, Universit\"at Wien\\
    Boltzmanngasse 5, A--1090 Wien, Austria 
\\[0.5cm]
    \footnotemark[3]Institut f\"ur Theoretische Physik, Technische
    Universit\"at M\"unchen,          \\
    James--Franck--Stra\ss e, D--85748 Garching, Germany
    \end{center} }
\vspace*{1cm}
\end{center}

{\Large \bf \begin{center} Abstract \end{center} }

The predictive power of top--condensation models strongly depends on the 
behaviour of higher dimensional operators. These are analyzed in this
paper by an extension of the standard renormalization group (RG) arguments 
which turns out to 
be a surprisingly powerful tool. Top--condensation models intermediated 
by underlying scalar exchange can be shown to be mere reparametrizations 
of the standard model. Further on, RG--arguments show that dynamical 
vector states cannot be lowered in top--condensation models. Finally we 
give a general argument concerning the size of higher dimensional 
operators of heavy vector exchange.

\renewcommand{\baselinestretch}{1.2}

\newpage
\renewcommand{\thefootnote}{\arabic{footnote}}
\setcounter{footnote}{0}


\section{Introduction}
\label{intro}

Dynamical symmetry breaking is an alternative of spontaneous electro--weak
symmetry breaking which replaces the fundamental Higgs field of the \SM by a
composite scalar field. Instead of fundamental scalars the simplest
model of top--condensation \cite{mty,bhl} introduces a new four--Fermion
interaction capable of forming the electro--weak symmetry breaking 
top--condensate\footnote{For an overview see e.g. ref. \cite{over}}. 
The dynamics then generates an effective scalar sector which describes the
symmetry breaking in analogy to the Ginzburg--Landau description of
superconductivity. 

It soon turned out that top--condensation requires a more specific picture of
the underlying interaction. The works of Hasenfratz et al. \cite{has} and 
Zinn-Justin \cite{zinn} showed that every standard model scenario 
can be reparametrized
as a top--condensation model if one has total freedom in the choice of higher
dimensional operators. 
To control these higher dimensional operators several specific models of 
underlying interactions were constructed \cite{fund,clague,topcolour}. 
In this work we analyze the structure of higher dimensional operators in the 
framework of the renormalization group (RG) approach. We
show, that a number of questions, which remained unsolved or only partially 
solved during the last years in the framework of a dynamical description, 
can be answered in an elegant way by a consequent use of RG--arguments.

Section 2 gives a short introduction to the RG--formulation of 
top--condensation including higher dimensional operators. This method was 
introduced in one of the founding papers of that model by Bardeen, 
Hill and Lindner \cite{bhl} to derive quantitative predictions of standard
top--condensation. Its consistency with the improved 
dynamical description of the Pagels--Stokar formula was shown in \cite{bdl}. 
Though the inclusion of higher dimensional operators was sketched e.g. 
in \cite{topcolour}, the theoretical implications of this generalized 
approach were never really exploited. To do so will be the goal of this work. 

Section 3 rephrases in a RG--framework the discussion about higher 
dimensional operators led in \cite{has} and \cite{zinn} which allows
a very intuitive understanding of that problematics. 
After a general discussion of relevant and irrelevant operators
from a RG--point of view in section 4, we have the tools to reach out to
several new statements about the structure of top--condensation models.
In section 5 it turns out that top condensation via heavy scalar exchange 
is an example of a pure \SM reparametrization. 
Section 6 shows that it is not possible 
to lower the scale of composite vector states in top--condensation models.
Section 7 gives a RG--argument against the existence of relevant higher 
dimensional operators stemming from heavy vector exchange.
After a comparison of our RG--arguments with dynamical methods in section 8
we draw our conclusions.


\section{Renormalization Group Approach}
\label{RGEA}

The simplest realization of the idea of top--condensation 
\cite{topcon,mty,bhl}, is a Nambu--Jona-Lasinio \cite{nambu} like model,
consisting of the kinetic parts of the ordinary quarks, leptons and
$SU(3)_c\times SU(2)_L\times U(1)_Y$ gauge fields and a new attractive
four--Fermion interaction. The Lagrangian is\footnote{We don't write colour
indices explicitly} 
\beq
{\cal L} = {\cal L}_{kin} + G\overline{\psi}_Lt_R\overline{t}_R\psi_L~,
\label{Lbhl}
\eeq
where ${\cal L}_{kin}$ contains the kinetic terms for all gauge fields,
quarks and leptons. $\psi^T_L=(t_L,b_L)$ is the third generation doublet of
quarks containing the left--handed top-- and bottom--fields and $t_R$ is the
right--handed component of the top--quark. Due to the non--renormalizable
structure of the model it is necessary to introduce a high energy cutoff
$\Lambda$. The model can then be studied in the large $N_c$ limit
(where $N_c$ is the number of colours).
The so derived gap equation 
\bea
m_t = \frac{1}{2}G<\bar{t}t> 
=\frac{N_cGm_t}{8\pi^4}\int\limits_0^{\Lambda} 
d^4k\frac{i}{k^2-m_t^2}
\label{gappi}
\eea
is found to be critical for $G> G_{cr} = 8\pi^2/N_c\Lambda^2$ and a
top--condensate emerges which leads to a top--mass  
$m_t = \frac{1}{2}G<\overline t t>$. A typical feature of top--condensation is 
the separation of the cutoff scale $\Lambda$ from the top--mass
scale. This is achieved by a fine--tuning of $G$ towards $G_{cr}$. 

The separation of scales allows a more elegant approach to calculate the
predictions of top--condensation, the so called 
renormalization group approach. 
To describe this concept we use the auxiliary field
formalism where the four--Fermion coupling $G$ is intermediated by a
non--propagating scalar doublet $\varphi$ of mass $G^{-1}$. This leads to the
Lagrangian
\bea
{\cal L} =
        {\cal L}_{kinetic}
        -\overline{\psi}_L \varphi t_R-\overline{t}_R \varphi^\dagger \psi_L
        -G^{-1} \varphi^\dagger \varphi \; .
\label{hfl}
\eea
At low energies a propagating Higgs field should emerge as a
top--antitop--boundstate. This  means that the Lagrangian of the \SM can be
seen as the low energy effective Lagrangian of the top--condensation model.
The \SM Lagrangian
\bea
{\cal L}_{SM} = {\cal L}_{kinetic}
          +\left( D_\mu\phi \right)^\dagger \left( D^\mu\phi \right)
          -\frac{\lambda}{2}  \left( \phi^\dagger\phi \right)^2
          +m^2 \phi^\dagger\phi
          -g_t \left( \overline{\psi}_L \phi t_R+\overline{t}_R \phi^\dagger
          \psi_L \right) 
\label{sml}
\eea
can be rewritten using $\varphi:=g_t\phi$ for a better comparison with
eq.~(\ref{hfl}):
\bea
{\cal L} = {\cal L}_{kinetic} +
          \frac{1}{g_t^2} \left( D_\mu\varphi \right)^\dagger
          \left( D^\mu\varphi \right)
          -\frac{\lambda}{2g_t^4}  \left( \varphi^\dagger\varphi \right)^2
          +\frac{m^2}{g_t^2} \varphi^\dagger\varphi
          -\left( \overline{\psi}_L \varphi t_R
          +\overline{t}_R \varphi^\dagger \psi_L \right)
          \label{lsm}
\eea
Now we have to require that eq.~(\ref{lsm}) becomes eq.~(\ref{hfl})
at a certain scale $\Lambda$. This leads to the following conditions:
\bea 
\lim_{\mu^2\to\Lambda^2}g_t^{-2}(\mu^2)= 0~,     \quad 
\lim_{\mu^2\to\Lambda^2}\frac{\lambda(\mu^2)}{g_t^4(\mu^2)}= 0~, \quad
\lim_{\mu^2\to\Lambda^2}\frac{m^2(\mu^2)}{g_t^2(\mu^2)}= -G^{-1}~,
\label{con}
\eea 
where $\Lambda$ is the high energy cutoff of the top--condensation
model. In underlying theories this cutoff corresponds to the mass of the heavy 
interaction particles. 

It is important to notice that the conditions of eq.~(\ref{con}) have
to obey both, renormalization group running and temperature--like quadratic
running\footnote{The name RG--approach therefore is a little bit misleading. We
nevertheless use this name as the name under which the concept was
introduced.}. 
The goal is to connect the low energy \SM with the tree level top--condensation
Lagrangian at the cutoff scale. First we reach the \SM at the cutoff scale 
using the renormalization group running of the \SM parameters. 
But this is not the whole story. The \SM is the effective theory of
top--condensation with all its cutoff regularized loop contributions.
These corrections are responsible for the symmetry breaking structure 
of the \SM scalar potential. To get 
the top--condensation tree level Lagrangian, it is necessary to subtract
the cutoff regularized loop contributions. 
However we describe the situation in the framework of the effective
theory, thus we need something which mirrors these subtractions 
in the effective theory. The quadratic running  of the Higgs mass parameter 
$m^2=\lambda v^2$ does this job. 

To see this explicitly we have
a closer look at the top-condensation mechanism at the level of loop summation
in the large  $N_c$--limit:
We start off with an effective top--condensation model with the
Lagrangian eq.~(\ref{Lbhl}). Now we write down the 4-fermion interaction term
including corrections in the large $N_c$--limit which comes up to the 
infinite bubble sum shown in fig.~\ref{bs4}.

\begin{minipage}{14.9cm}
\begin{center}
\begin{picture}(340,60)(0,0)
\ArrowLine(0,10)(20,30)         \ArrowLine(20,30)(0,50)
\Vertex(20,30) {1.5}
\ArrowLine(20,30)(40,10)        \ArrowLine(40,50)(20,30)
\ArrowLine(80,10)(100,30)       \ArrowLine(100,30)(80,50)
\Vertex(100,30) {1.5}
\ArrowArc(120,30)(20,0,180)
\ArrowArc(120,30)(20,180,0)
\Vertex(140,30) {1.5}
\ArrowLine(140,30)(160,10)      \ArrowLine(160,50)(140,30)
\ArrowLine(200,10)(220,30)      \ArrowLine(220,30)(200,50)
\Vertex(220,30) {1.5}
\ArrowArc(240,30)(20,0,180)
\ArrowArc(240,30)(20,180,0)
\Vertex(260,30) {1.5}
\ArrowArc(280,30)(20,0,180)
\ArrowArc(280,30)(20,180,0)
\Vertex(300,30) {1.5}
\ArrowLine(300,30)(320,10)      \ArrowLine(320,50)(300,30)
\Text(60,30)[c] {$+$}         
\Text(180,30)[c] {$+$}
\Text(340,30)[l] {$+ \; \cdots$}
\end{picture}
\end{center}
\end{minipage}
\begin{fig}
\begin{center}
Bubble sum for the (pseudo)scalar boundstate propagator
\label{bs4}
\end{center}
\end{fig}

This summation leads to an amplitude
\beq
\Gamma(p^2) = \frac{1/N_c}{1/G - I},
\label{prop}
\eeq
where $p$ is the outer momentum of the quarks and
$G^2I$ is the integral over one bubble up to a cutoff $\Lambda$. There
are four different amplitudes of the type eq.~(\ref{prop}). A scalar 
and a pseudoscalar for {\it top--top} and positively and negatively charged
pseudoscalar {\it top--bottom} amplitudes. To see the principle we just look 
at the scalar {\it top--top} amplitude. We get
\bea
I_{s4}(p^2)=\frac{-i}{2}tr\int \frac{d^4k}{(2\pi )^4}
\frac{(\slash{k}+m_t)(\slash{k}-\slash{p}+m_t)}{(k^2-m_t^2)[(k-p)^2-m_t^2)} 
\eea
which can be written as
\bea
I_{s4}(p^2)=\underbrace{\frac{1}{8\pi^4}\int d^4k\frac{i}{k^2-m_t^2}}_{I_1}-
\underbrace{\frac{(p^2-4m_t^2)}{8\pi^4}\int d^4k\frac{i}{(k^2-m_t^2)
[(k-p)^2-m_t^2]}}_{I_2}
\label{prope}
\eea
where we have separated a quadratically divergent term $I_1$ and a 
logarithmically divergent term $I_2$. However $I_1$ is exactly the tadpole
of the gap equation eq.~(\ref{gappi}) so that $1-GI_1=0$. Inserting this into  
eq.~(\ref{prop}) we get an amplitude 
\bea
\Gamma_{s4}(p^2)=\frac{1}{N_cI_2}=\frac{1}{(p^2-4m_t^2)
            N_c\frac{1}{(4\pi)^2}\ln\frac{M^2}{p^2}+\mbox{finite terms}}
\eea
where all large couplings $G$ have fallen out. The use of the finetuned
gap--equation led to a scalar propagator with a small mass 
$M_H=2m_t$.\footnote{The mass relation $M_H=2m_t$ looks very intuitive for a
top--antitop boundstate but is only valid in this simple approximation.
The RG--approach gives a lower value.} This scalar top--antitop boundstate 
shows up as an effective Higgs field in the low energy 
Lagrangian.\footnote{the pseudoscalar {\it top--top} and {\it top--bottom}
amplitudes lead to the massless Goldstone bosons following the same arguments.}

Now we introduce an infrared cutoff $\mu$ in the loop integrals and shift
that cutoff continuously up towards $\Lambda$. By doing so we
reduce the loop contributions  until we finally get back to
tree--level top--condensation for $\mu=\Lambda$. 
After shifting $\mu$ above the electro--weak scale we loose the relation 
$1-GI_1=0$ and therefore the finetuning.
Thus we can understand eq.~(\ref{prope}) approximately as a heavy scalar 
propagator of mass $\mu$. At $\mu=\Lambda$ all loop contributions have vanished
and we end up with a nonpropagating auxiliary Higgs like it was introduced 
in eq.~(\ref{hfl}). If we write our Higgs propagator into the 
effective Lagrangian, we see explicitly now, that a continuous subtraction
of the loop contributions appears there as a quadratic running of the 
scalar mass term. This makes it possible to meet the third 
condition of eq.~(\ref{con}).\footnote{For the sake of simplicity we have 
developed our argument in the framework of simple loop summation. To get 
an exact matching between the logarithmic contributions of the dynamical 
calculations and the RG--calculations, it would be necessary to use an
improved approach involving a running top mass function. (See~\cite{bdl})
However, the simpler approach is sufficient to understand the role of the
quadratic running.}

The fulfillment of the three conditions by the described running implies
certain values for the top-- and the Higgs mass depending on the VEV
(fixed by the W--mass) and
the scale of the Landau pole of the top--Yukawa coupling (fixed by the 
cutoff scale). The fulfillment of the pole conditions by RG--running compared
with different \SM scenarios (fig.~2) is shown in fig.~3 for a cutoff
scale of $10^{4}$ GeV. 

These conditions give a too high top--mass prediction in the simplest model.
However one can construct extensions by introducing additional four--Fermion 
couplings 
\cite{ext2h}, enlarging the gauge group \cite{extlr} or supersymmetrizing 
the theory 
\cite{extsusy} to achieve a phenomenologically viable value.

Up to now we have described the RG--approach in the framework of minimal
top--condensation. Now we want to ask what are the general conditions which
make a RG--description of a theory of dynamical symmetry breaking possible.

\begin{figure}[b]
\hspace*{0.25cm}
\bild{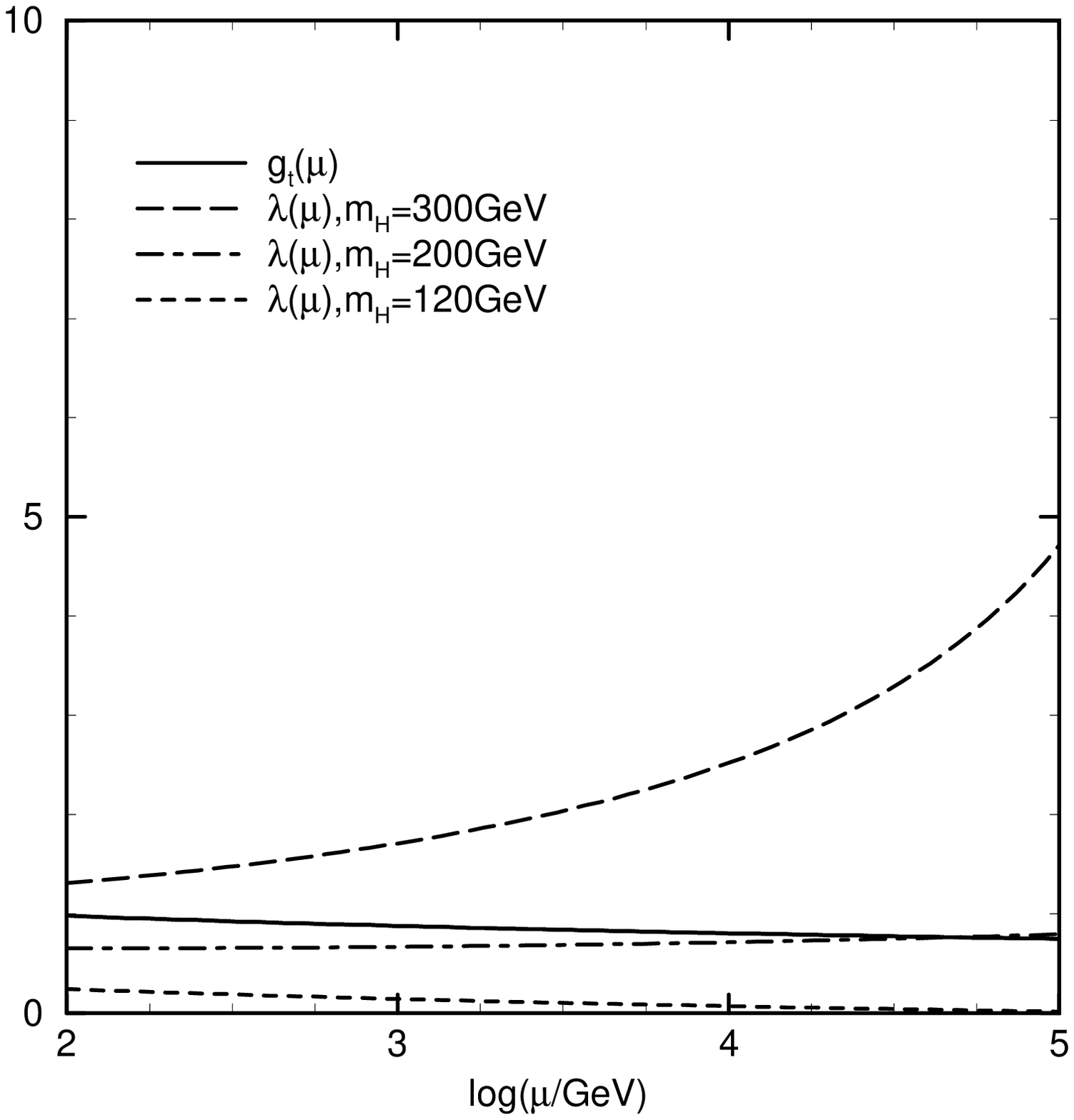}
     {\Text(60,150)[c]{}
      \Text(30,75)[c]{} }
     {RG--evolution in the standard model for $\left(m_t\right)_{\overline{MS}}
      = 167\;GeV$.}  
\hspace*{0.4cm}
\bild{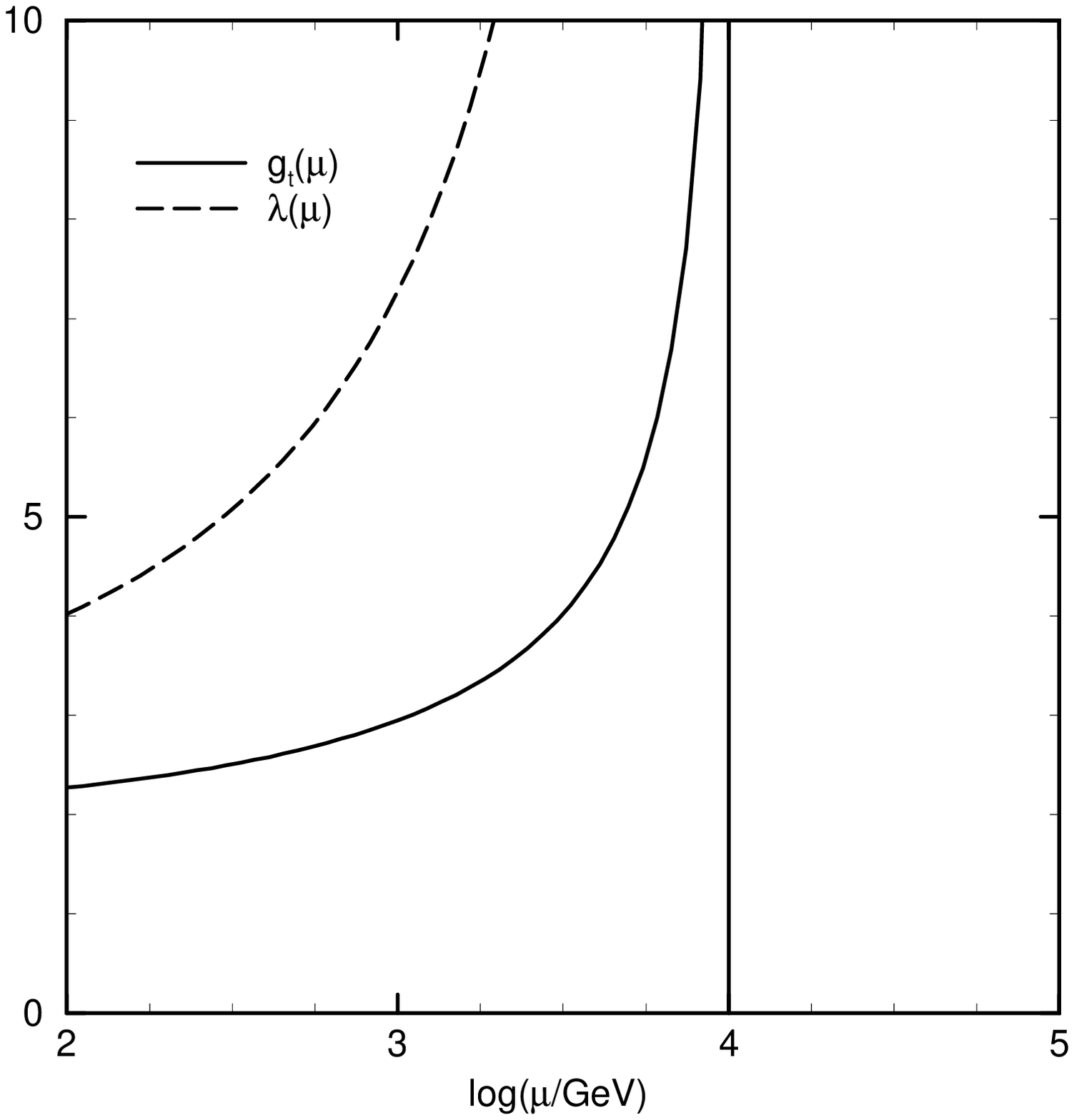}
     {\Text(60,150)[c]{}
      \Text(30,75)[c]{} }
     {RG--evolution in the simple top--condensation model for a cutoff scale 
      $\Lambda = 10^{4}\;GeV$.} 
\hspace*{0.25cm}
\end{figure}
\begin{figure}[b]
\hspace*{0.25cm}
\bild{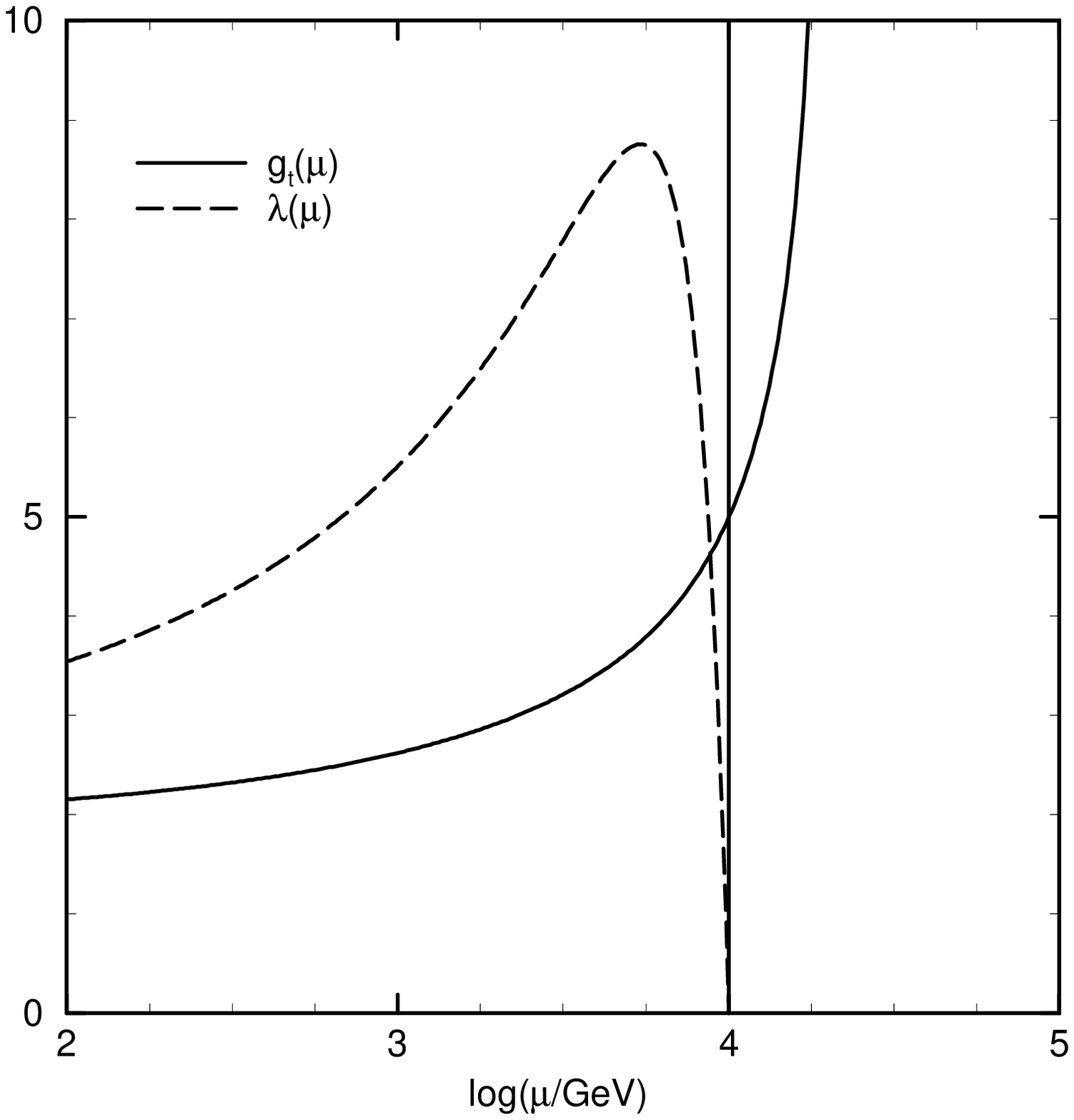}
     {\Text(60,150)[c]{}
      \Text(30,75)[c]{} }
     {RG--evolution for the case of heavy scalar exchange.} 
\hspace*{0.4cm}
\bild{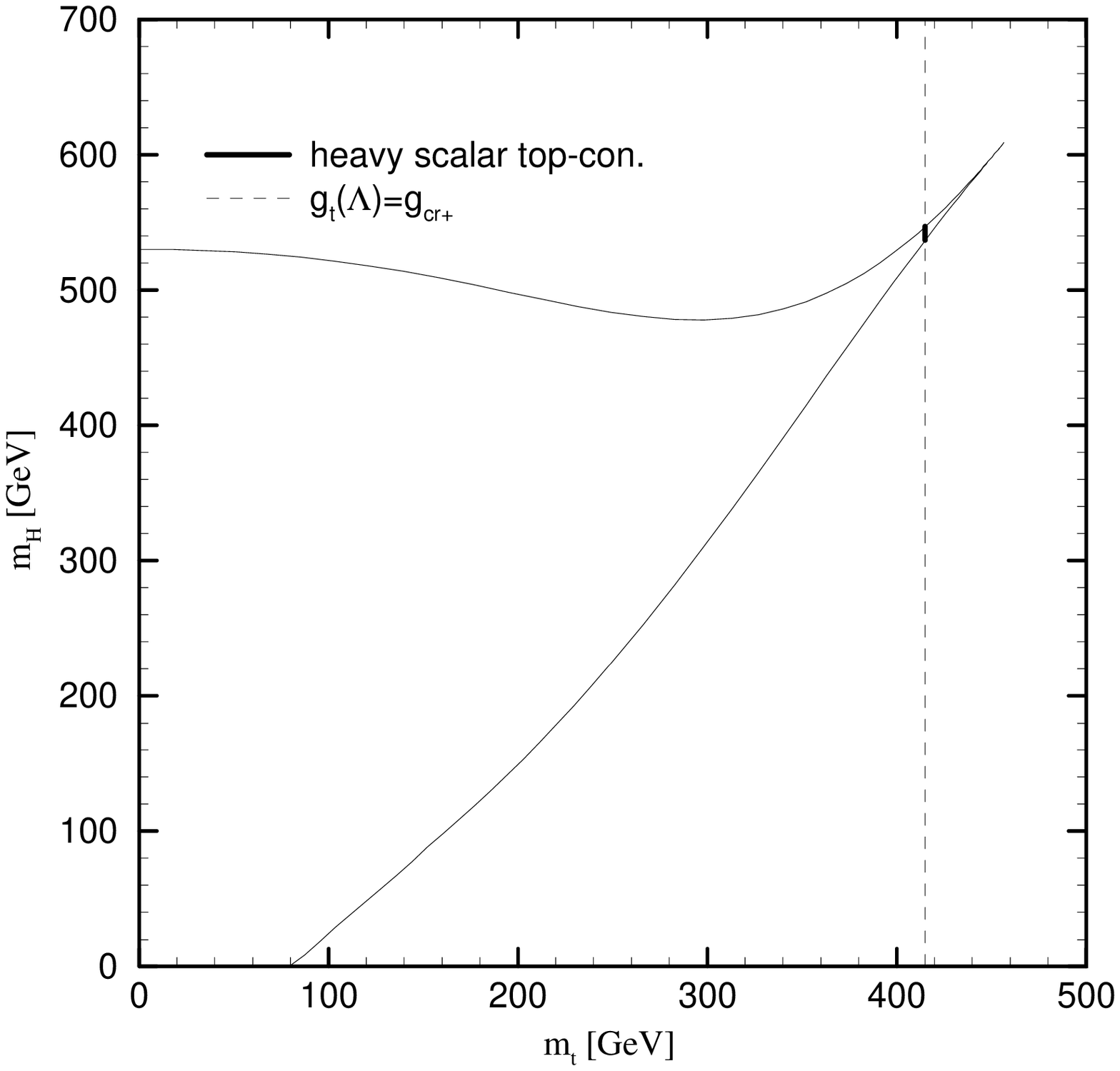}
     {\Text(65,100)[c]{\scriptsize allowed parameter space}
      \Text(65,85)[c]{\scriptsize in the standard model} }
     {Allowed parameter space in the \SM and in ``heavy scalar mediated
      top--condensation'' for $\Lambda = 10^4\;GeV$.} 
\hspace*{0.25cm}
\end{figure}

Two preconditions have to be met: 
First the scale of the interaction responsible for dynamical symmetry breaking
must be considerably higher
than the mass scale of the bound states. If this separation of scales
does not exist the region where a RG--description could apply is zero.
Second the light or 
massless fields of the full theory must also appear in the low energy theory.
This is necessary to make an identification of parameters of the effective
and the full theory possible.
The first condition is not fulfilled in technicolour models where the new
interaction scale is not separated from mass scale of the techniquarks.
The second condition is neither fulfilled in technicolour nor 
in preonic models because techniquarks and preons
do not appear in the low energy theory due to confinement.
If however both conditions are fulfilled, as it is the case in 
top--condensation, a RG--formulation always exists and is not devaluated by 
higher dimensional operators.  One can always formulate 
a low energy  effective theory which includes the whole dynamics of the model.
From this low energy effective theory one must get back to the tree structure 
of the full theory if one considers RG--running up to the critical scale and 
subtracts all loop corrections. We have seen in the minimal case that this 
subtraction corresponds to considering quadratic running of the scalar mass 
parameter in the effective theory. 
This situation is not changed at all by higher dimensional operators. 
The higher dimensional operators just change the identification conditions
for the effective theory which are based on renormalization group running
\cite{topcolour}.

In the forthcoming sections we will use this extension of the RG--approach
to learn about models with higher dimensional operators.
Our basic idea will be the following: We will
assume that a RG--formulation of the discussed model exists. Then we 
investigate what the implications of the mere existence of such a 
formulation are. It will turn out that these implications are by far
stronger than one would expect.
For sake of simplicity we do our discussions in the simplest model. The 
conclusions apply to all non--supersymmetric generalizations
\footnote{In a RG--treatment of supersymmetric top--condensation
quadratic running plays no role due to the cancellations of 
quadratic divergencies. Therefore the arguments of our discussion are 
not valid.}.


\section{Higher Dimensional Operators}
\label{HDO}

Soon after the invention of top--condensation Hasenfratz et al.\cite{has}
and Zinn-Justin \cite{zinn} showed that top--condensation with a specific
set of higher dimensional operators represents a reparame\-tri\-zation of
the standard model. We will reproduce their arguments in the framework of
the extended RG--approach which gives a very intuitive picture.
We consider a low energy \SM scenario with arbitrary top-- and Higgs mass.
Now we just follow the procedure described at the end of the last section.
We choose an arbitrary cutoff scale and ask, whether it is possible to connect
the low energy \SM Lagrangian with a general top--condensation Lagrangian at
this cutoff scale. As it will be shown in the following the answer to this 
question is yes.
The renormalization group running in general will not lead to a Landau pole
at the cutoff scale, we will still face scalar kinetic terms there. The
quadratic running of the Higgs mass parameter will give a positive mass
term of the order $\Lambda$  just like in standard top--condensation. 
To match the two Lagrangians we therefore have to find a correspondence 
to the kinetic terms of the \SM Higgs field in the top--condensation 
Lagrangian. We start with the low energy Lagrangian
eq.~(\ref{sml}) neglecting the gauge couplings for
simplicity.
We consider renormalization group running up to the cutoff scale $\Lambda$
which leads to the Lagrangian
\beq
{\cal L}_{SM}(\Lambda) = {\cal L}_{kin}(\Lambda)
          +\left( \partial_\mu\phi \right)^\dagger \left( \partial^\mu\phi
          \right)
          -\frac{\lambda(\Lambda)}{2}  \left( \phi^\dagger\phi \right)^2
          +m^2(\Lambda) \phi^\dagger\phi
          -g_t(\Lambda) \left( \overline{\psi}_L \phi 
          t_R+\overline{t}_R \phi^\dagger \psi_L \right)~.
\label{hesl} 
\eeq
We consider quadratic running (qr) of the scalar mass term which leads to
\beq
{\cal L}_{qr}(\Lambda) = {\cal L}_{kin}(\Lambda)
          +\left( \partial_\mu\phi \right)^\dagger \left( \partial^\mu\phi
          \right)
          -\frac{\lambda(\Lambda)}{2}  \left( \phi^\dagger\phi \right)^2
          -\frac{g_t^2(\Lambda)}{g_{cr_+}^2}\Lambda^2 \phi^\dagger\phi
          -g_t(\Lambda) \left( \overline{\psi}_L \phi t_R+\overline{t}_R 
           \phi^\dagger \psi_L \right)~,
\label{heql} 
\eeq
where we defined a dimensionless coupling $g_{cr_+}$ by 
$G:=\frac{g_{cr_+}^2}{\Lambda^2}$. In a fine--tuned model ($G \cong G_{cr}$)
$g_{cr_+}$ is only slightly above the critical value $g_{cr}$, the corrections 
are of the order $g_{cr_+}=g_{cr}+ O(\frac{\mu_{elw}^2}{\Lambda^2}
ln\frac{\Lambda}{\mu_{elw}})$. 
We can see that, as we keep the terms  
$\left( \partial_\mu\phi \right)^\dagger\left( \partial^\mu\phi \right)$
at the cutoff scale, the redefinition $\varphi :=g_t\phi$
is not infinite. However, to get a clear picture of what is going on, we
choose a redefinition $\varphi:= \frac{g_t(\Lambda)}{g_{cr_+}}\phi$. We get
\beq
{\cal L}_{qr}(\Lambda) = {\cal L}_{kin}(\Lambda)
       +\frac{g_{cr_+}^2}{g_t^2(\Lambda)}
       \left( \partial_\mu\varphi \right)^\dagger \left( \partial^\mu\varphi
       \right)
       -\frac{\lambda(\Lambda)g_{cr_+}^4}{2g_t^4(\Lambda)}  
       \left( \varphi^\dagger\varphi \right)^2
       -\Lambda^2 \varphi^\dagger\varphi
       -g_{cr_+} \left( \overline{\psi}_L \varphi 
        t_R+\overline{t}_R \varphi^\dagger \psi_L \right)~.
\label{var} 
\eeq
We have separated the cutoff suppression scale from the Yukawa coupling of the
auxiliary field. 
The heavy mass term represents
purely the compensation of the cutoff contribution while the Yukawa 
coupling controls dynamical symmetry breaking and is fine--tuned against 
the coupling where the perturbation expansion breaks down.
We are very close to a scenario with heavy scalar exchange of mass 
$\Lambda$ in a strong coupling regime.
The only deviation is the factor $\frac{g_{cr_+}^2}{g_t^2(\Lambda)}$
in front of the scalar kinetic term. Of course one could find a different
normalization for $\varphi$ where this factor is one, but in this case we 
would loose the connection between the heavy mass and the cutoff scale as well
as the right size for the Yukawa coupling which has to be strong to induce 
dynamical symmetry breaking, if we take the interpretation of our Lagrangian as
a description of heavy particle exchange seriously. What we have
constructed is therefore not heavy scalar exchange but just 
a top--condensation 
model with higher dimensional operators in auxiliary field formalism. 

We expand now in the heavy scale ${\Lambda}$ by inserting the Euler--Lagrange
equations for $\varphi$ and $\varphi^\dagger$ repeatedly to get a series 
of higher dimensional operators. The Euler Lagrange equations are
\bea
\varphi &=& \frac{1}{\Lambda^2}\left[-g_{cr_+}\overline{t}_R\psi_L-
\frac{g_{cr_+}^2}{g_t^2(\Lambda)}\partial^2\varphi-
\frac{g_{cr_+}^4\lambda(\Lambda)}{g_t^4(\Lambda)}
\varphi~\varphi^\dagger\varphi\right]~,
\label{el1} \\
\varphi^\dagger &=& \frac{1}{\Lambda^2}\left[-g_{cr_+}\overline{\psi}_L t_R-
\frac{g_{cr_+}^2}{g_t^2(\Lambda)}\partial^2\varphi^\dagger -
\frac{g_{cr_+}^4\lambda(\Lambda)}{g_t^4(\Lambda)}
\varphi^\dagger\varphi~\varphi^\dagger\right]~,
\label{el2} 
\eea
which leads to the following expansion of  $\varphi$ and $\varphi^\dagger$:
\beq
\varphi=-\frac{g_{cr_+}}{\Lambda^2}\overline{t}_R\psi_L
+\frac{g_{cr_+}}{{\Lambda}^4}\partial^2
\Big(\overline{\psi}_L t_R\Big) + O(\Lambda^{-6})~,
\label{phi}
\eeq
\beq
\varphi^\dagger=-\frac{g_{cr_+}}{{\Lambda}^2}\overline{\psi}_L 
t_R +\frac{g_{cr_+}}{{\Lambda}^4}\partial^2
\Big(\overline{t}_R\psi_L\Big)+ O({\Lambda}^{-6})~.
\label{phi+} 
\eeq
Reinserting this expansion into eq.~(\ref{var}) leads to the new 
top--condensation Lagrangian:
\beq
{\cal L}_{tc}(\Lambda) = 
\frac{g_{cr_+}^2}{{\Lambda}^2}\overline{\psi}_Lt_R\overline{t}_R\psi-
\frac{g_{cr_+}^2}{\Lambda^4}\Big(\partial_{\mu}\Big(\overline{t}_R
\psi_L\Big)\Big)^\dagger \Big(\partial_{\mu}\Big(\overline{t}_R
\psi_L\Big)\Big)+O({\Lambda}^{-6})
\label{hdtcl} 
\eeq
The resulting higher dimensional operators are those suggested by
Hasenfratz et al. This discussion tells us that
every \SM Lagrangian can be understood as the low energy effective theory of a
top--condensation model with suitable higher dimensional operators. 
There exists a specific set of higher dimensional operators for any
choice of the top condensation cutoff scale below eventual Landau poles. 

This does not automatically mean that top--condensation models including
such sets of higher dimensional operators are meaningless. If some couplings
become non--local at the
cutoff scale, the top--condensation could still be distinguished from the 
standard model. It would be a reasonable theory in its own
right, though not easily detectable at low energies. 
If however such nonlocality does not appear at the cutoff scale, 
the higher dimensional operators just describe the propagation 
of the Higgs at this scale. In this case the \SM is {\bf not} just a 
low energy effective theory of top--condensation up to the cutoff 
but top--condensation and the \SM are identical at all scales.
They are just different parametrizations of the same model.


\section{Relevant and Irrelevant Operators}
\label{IRL}

In order to get a clear understanding which higher dimensional operators
turn a specific model of top--condensation into a new theory and which 
ones keep it a pure reparametrization of the
\SM we have to introduce the notations of relevant and irrelevant operators in
the renormalization group framework.
Relevant operators at the electro--weak scale are those operators which
contribute there without a high suppression scale. Usually all operators
stemming from the exchange of heavy particles or from non--renormalizable
interaction at a high scale will be irrelevant at low scales. This is however
not always the case. Especially in top--condensation, being a fine--tuned
theory, the relevance of higher dimensional operators suppressed
by the cutoff scale is a basic principle of the theory. For example
the four--Fermion coupling term
\beq
O_4=G\overline{\psi}_Lt_R\overline{t}_R\psi_L
\label{4f}
\eeq
is suppressed by the heavy scale. However if it would not
be relevant at the electro--weak scale, we could not see any 
light electro--weak breaking phenomenology induced by top--condensation.
The relevance of this operator can be seen in the framework of the
RG--approach. In auxiliary fields eq.~(\ref{4f}) appears as a heavy mass
term
\beq
O_{4H}=G^{-1}H^\dagger H~,
\eeq
where $H=G\overline{\psi}_Lt_R$. Now the quadratic running connects this term 
to the small Higgs mass parameter of the \SM which obviously is relevant 
at low scales. We see that the fine--tuning of the theory has made highly
suppressed operators relevant at low scales, which is formally described 
by the quadratic running of the scalar mass parameter from a high to a 
low value. There is no reason why
the simple four--Fermion coupling of eq.~(\ref{4f}) 
should be the only operator in a 
top--condensation model that is relevant at the electro--weak scale. We have
seen other such operators already in section \ref{HDO},
where the higher dimensional operators 
obviously change the low energy phenomenology considerably and therefore
must be relevant. On the other side not every suppressed 
operator in a top--condensation model is relevant at low scales. 
Our goal is now to 
find a way to distinguish between relevant and irrelevant operators.

It is possible to give a simple general definition of 
a relevant operator in the framework of the renormalization group approach. 
We start with some model at a high scale $\mu_h$ consisting
of a number of mass terms, dimension--four operators and higher 
dimensional operators. This model must have an effective theory 
at a lower scale $\mu_l$ in which all higher dimensional operators 
suppressed by $\mu_h$ are neglected. The renormalization group
approach connects the low energy effective 
theory with the full theory at $\mu_h$ by RG-- and quadratic
running. In general, however, not all operators of the full theory 
can be identified with operators of the effective theory. 
Those which can be identified are low energy relevant operators as
they contribute to the low energy effective theory in form of the
identifiable effective operators. Those which cannot be identified
are irrelevant. They occur at low energies only in form of operators 
suppressed by the high scale and can be neglected.

Using the notation of irrelevant operators it is possible to give a
clear definition of what is a mere reparametrization of the \SM and
what is a theory in its own right:

A theory that is distinguishable from the \SM must have irrelevant operators
at the high scale. These irrelevant operators are not part of the 
effective theory and therefore distinguish this effective theory from 
the full theory at higher scales. A theory which does not have irrelevant
operators is fully described by the ``effective'' theory and therefore
not more than its reparametrization. We will see examples for both cases
in the upcoming sections. 
The simplest example for a mere reparametrization is minimal top--condensation
without any underlying concept, where all high scale operators are relevant
at low energies.\footnote{ This case is a little bit special however. 
While the \SM cannot be defined above the Landau pole top--condensation
formally could due to the infinite
field redefinition in course of the RG--approach. 
However this cannot be more than a formal continuation
since the notation of a dimensionful non--renormalizable coupling is not
physically senseful above that coupling scale.}
To give top--condensation a physical meaning it is necessary to introduce 
an underlying concept that contains irrelevant operators.
The next section will discuss that heavy scalar exchange is no viable 
possibility since it is a standard model reparametrization itself.
An example for a true underlying theory in its 
own right is heavy vector exchange which we will discuss in section \ref{hve}.


\section{Heavy Scalar Exchange}
\label{hase}

Our method in section \ref{HDO} was to construct exactly the set of higher
dimensional operators which can be connected with a specific standard model 
Lagrangian
through RG-- plus quadratic running. Due to the discussion in section 4 
this means that we do not have any irrelevant operators in our Lagrangians. 
Therefore we constructed in fact pure reparametrizations of the standard
model.

The higher dimensional operators of section \ref{HDO} however look 
very similar to those produced by heavy scalar exchange.
Actually they are identical with heavy scalar exchange for the choice
$g_t(\Lambda)=g_{cr_+}$.
In this case the factor in front of the scalar kinetic term of 
eq.~(\ref{var}) is one and we get pure heavy scalar mediated top--condensation.
Heavy scalar exchange thus does not produce any irrelevant operators,
it is not a theory in its own right but just a special case of the 
standard model reparametrizations constructed in section \ref{HDO}. It is a 
reparametrization of a \SM with a Yukawa coupling $g_t$ that becomes 
strong at the mass scale of the heavy scalar.

There have been investigations of scalar mediated top--condensation in
\cite{clague} where the results differ
from the \SM case. We will just argue shortly why these results are incorrect 
according to our understanding.

\cite{clague} uses the standard model scalar sector, however with a positive
heavy mass term in the scalar potential:
\beq
   V(\Phi)=m_{\phi}^2|\Phi|^2+\lambda|\Phi|^4
\eeq
Only the s-channel scalar exchange diagrams contribute in lowest order 
$1/N_c$. 
For a sufficiently large top--Yukawa coupling top--condensation can occur.
The authors observe that the top--condensate gives an additional contribution
to the scalar potential which becomes
\beq
   V(\Phi)=m_{\Phi}^2|\Phi|^2+\lambda|\Phi|^4+
   \frac{g_t}{\sqrt{2}}<\overline{t}t>\phi_3
\eeq 
and produces a final VEV of $\phi_3$ which is
\beq
   <\phi_3>=v=-\frac{g_t<\overline{t}t>}{\sqrt{2}m_\phi^2}~.
   \label{repr}
\eeq 
This is interpreted as a sign for a light component of the fundamental 
scalar field in addition to the composite Higgs field. However this is  
not correct. The misinterpretation is based on a
misunderstanding of the role of the quadratic 
contributions in top--condensation. As discussed in section \ref{RGEA} 
these connect the tree level top--condensation Lagrangian with its effective
low energy Lagrangian. Therefore the light scalar component found above
after considering quadratic loop contributions cannot be interpreted in 
the framework of top--condensation, it has to be understood in the framework 
of the effective theory. With this knowledge the message of eq.~(\ref{repr})
is very clear. The standard model scalar can be identified with the heavy 
scalar of top--condensation corrected by its quadratic contributions. What 
is found in \cite{clague} is not an additional scalar but nothing else than
the standard model Higgs. 

Figures 2--5 illustrate the three scenarios discussed so far. Fig.~2 
describes the RG--running of the top Yukawa coupling and the four Higgs
coupling in the \SM case with realistic parameters. It would be possible 
to use any scale below the Landau pole of the Higgs quartic coupling as a 
cutoff scale in an
extended top--condensation scenario as described in section \ref{HDO}.
Fig.~3 describes simple top--condensation without higher dimensional 
operators. Fig.~4 shows one case of heavy scalar exchange
($\lambda(\Lambda)=0$)
which however itself is just a special case of a \SM reparametrization.
One can see that the heavy scalar exchange for arbitrary four--scalar coupling 
$\lambda_S$ of the heavy scalar
always remains very close to the minimal top--condensation
case. The deviations of the low energy parameters from the minimal case
are a few percent.   This is a consequence of the so called quasi 
infrared fixed point behaviour of the RG--running that forces the parameters
into a small region of low energy values if they run down from  
Landau poles at considerably varying high scales. Nevertheless the operators
responsible for the difference between minimal and scalar intermediated
top--condensation of course must be classified as relevant operators for low 
energy physics. While the quasi fixed point for a high scale
around $10^{15} GeV$ suppresses the importance of high energy variations
to a few percent, irrelevant operator according to
our definition would be suppressed by $\mu_h/\mu_l \sim 10^{-13}$.

Fig.~\ref{plot_lin} \cite{lindner} shows the segment of the \SM parameter 
space which can be reparametrized by ``heavy scalar mediated 
top--condensation''. It is defined by the 
intersection of the allowed parameter region for the \SM case and the $m_t$
line which leads to the condition $g_t(\Lambda) = g_{cr_+}$.
The Higgs mass is not fixed exactly in this scenario, it depends 
on the value of the four--scalar coupling of the heavy scalars.
However due to the infrared fixed point behaviour the region of allowed 
Higgs masses, restricted by  the pole for $\lambda$  at the cutoff from 
above and by $\lambda=0$ at the cutoff from below is very small. The point in
parameter space corresponding to standard top--condensation would be the
intersection between vacuum stability and triviality bound (the lower and upper
bound of the allowed parameter space for the standard model). 

A last point which should be addressed is the t-channel scalar exchange.
Because of the $1/N_c$--expansion it is only the dominant case for 
a colour octet scalar interaction which is repulsive.


\section{Composite Vectors} 
\label{cv}

One open question in top--condensation has been the behaviour of 
vector bound states. In the simplest top--condensation model they do not
appear due to the specific structure of the four--Fermion coupling $G$.
However any natural extension of $G$, especially any effective coupling
stemming from heavy vector exchange, will include operators of the type
\bea
 G^{\prime}      \Big( \overline{\psi}_L\gamma_\mu{\psi}_L \Big) 
                 \Big(  \overline{\psi}_L\gamma^\mu{\psi}_L \Big) +
G^{\prime\prime} \Big( \overline{t}_R\gamma_\mu t_R \Big) 
                 \Big( \overline{t}_R\gamma^\mu t_R \Big) 
\label{veo}
\eea
and may  consequently produce vector boundstates. The question is, on which
scale these boundstates will appear in a fine--tuned model. Will they be
lowered like the scalar boundstates, will they remain at the heavy scale or
will they be somewhere in between?  Most of the investigations about vector
bound states in NJL--like models were done in the context of QCD (see for
example \cite{weise} and references therein) which is no fine--tuned model.
The question of scale for the vector bound states therefore does not arise in 
that case. In a top--condensation framework just the role of hypothetical 
vector boundstates at arbitrary scales was investigated in \cite{luest}. 
The actual scale of these fields remained unknown. 

It turns out that the extended RG--approach gives an answer to this question.  
If we imagine a vectorial boundstate with a mass considerably lower than the 
cutoff scale, then the RG--approach demands the identification of the 
corresponding operators of the effective theory with operators of the 
top--condensation theory at the cutoff scale. 
Since all top--condensation operators are connected to the cutoff scale 
the expansion in the heavy scale starts with dimension six operators: 
\bea
   {\cal L}_{tc} = \frac{1}{\Lambda^2}O(6)+\frac{1}{\Lambda^4}O(8)+\ldots \;\;.
\eea
Therefore the vectorial mass must be identified with a cutoff scale mass term. 
Due to the lower mass scale 
of the vector state this identification cannot be provided by RG--running. 
It must be provided by quadratic running. 
 Now our theory is a theory of 
dynamically broken gauge symmetry. The gap--equation
produces a condensate which breaks the symmetry and serves as a mass term 
for the light fields. The low scale of those masses is achieved by
a fine--tuning of the effective coupling $G$ 
towards the critical coupling of the gap equation.
Therefore any low mass term must be produced by the gap equation,
in other words any field which gets a low scale mass
has to couple to the condensate. 
In the effective theory the top--condensate
corresponds to the VEV of the Higgs field. Thus if vector boundstates
would acquire fine--tuned masses from the gap equation, these
would have to be produced by coupling the
fields to a Higgs VEV:
\bea
   {\cal L}_V = (\partial V)^2+V^2\!<\!v\!>^2 + \, \overline{\psi}
   \gamma^\mu \psi V_\mu
\eea
Of course fundamental mass terms for vector--boundstates can exist,
those are however not induced by the condensate but directly by the heavy 
interaction scale.
They have consequently no connection to the fine--tuning of the theory 
and therefore necessarily remain at the high scale.
The only possible low scale
quadratic mass terms in the effective theory are the mass parameters of 
the scalar potential that will finally produce the VEV. These mass parameters
however change their sign in course of their quadratic running at the 
breaking scale $\mu_{br}<\Lambda$, which puts the VEV to zero. 
Consequently all 
vector mass terms vanish at $\mu_{br}$ if we consider quadratic running. 
\bea
   {\cal L}_V^{qr}(\mu_{br}) = 
   (\partial V)^2 + \overline{\psi}\gamma^\mu\psi V_\mu
   \label{nom}
\eea
It is therefore not possible for a vector mass term to run up to 
a high scale by quadratic running. Thus one cannot connect
low vectorial mass terms with high scale operators of top--condensation.
All low scale vectorial mass terms remain low mass terms in the  
top--condensation theory, none of them can possibly be interpreted
as a dynamically produced boundstate.
All vector boundstates in a top--condensation theory must 
have a mass of the order cutoff scale.\footnote{The statement that 
vector--boundstates cannot be lowered by fine--tuning
seems to be rather general. For example it remains true for a 
Nambu--Jona-Lasinio model without dynamical
gauge symmetry breaking. The breakdown of the chiral symmetry due to
the Fermion condensate is sufficient to keep our argument valid.}


\section{Heavy Vector Exchange}
\label{hve}

Heavy vector exchange is the standard concept to introduce an underlying
theory for the non--renormalizable four--Fermion interaction. Of course
this concept implies additional higher dimensional operators at the cutoff
scale. Now once again there arises the question whether these operators
can change the low energy prediction of top--condensation. This question
was addressed in 
\cite{topcolour} where box diagrams of the type 

\begin{minipage}{14.9cm}
\begin{center}
\begin{picture}(120,80)(0,0)
\ArrowLine(20,75)(40,60)        \ArrowLine(40,20)(20,5)
\Vertex(40,60) {1.5}            \Vertex(40,20) {1.5}
\ArrowLine(40,60)(80,60)        \ArrowLine(80,20)(40,20)
\Vertex(80,60) {1.5}            \Vertex(80,20) {1.5}
\ArrowLine(80,60)(100,75)       \ArrowLine(100,5)(80,20)
\Photon(40,20)(40,60) 2 5 
\Photon(80,20)(80,60) 2 5
\Text(18,75)[r] {$p_1$}         \Text(18,5)[r] {$p_2$}
\Text(102,75)[l] {$p_1-q$}      \Text(102,5)[l] {$p_2+q$} 
\end{picture}
\end{center}
\end{minipage}
\parbox{1.5cm}{\beq \eeq}

\vspace{0.5cm}

are studied for a special example (a specific gauge structure).
The conclusion was that the scalar kinetic
contributions produced by these diagrams are about 100 times smaller
than it would be necessary to give a considerable effect.
This was however just shown for one example and restricted to box diagrams.
The statement that the situation will not change in other models or at
higher orders basically remained a plausibility argument.
We use the arguments developed in the last sections to get
a more substantial understanding of this question. 

The crucial point of this argument has already been stated in 
section \ref{RGEA}. The RG--approach connects the
tree level \SM Lagrangian (which is the effective formulation of
the top--condensation model with all corrections) 
to the tree--level top--condensation Lagrangian at the cutoff scale. 
We once more want to emphasize that this is not just the case in the 
minimal model. It is a necessary feature of all extensions as well, rooted in
the nature of fine--tuned dynamical symmetry breaking: There exist highly
suppressed tree level operators (in effective top--condensation 
four fermion coupling terms, in a full model e.g. strongly  
coupled heavy vector exchange), which nevertheless produce
low energy relevant effective operators via their dynamics 
(loop contributions). The mechanism of the RG--approach connects
these low energy relevant tree operators to their low energy remnants,
the operators of the standard model.  As the 
whole low energy phenomenology stems from the dynamics induced by the
tree level top--condensation Lagrangian, in the RG--framework one can say 
everything about the low energy 
theory by connecting it to this tree level Lagrangian. 
The fact that we connect to tree level operators
is reflected in the quadratic running of the scalar mass term which
represents the subtraction of the loop amplitudes enhanced by finetuning.

If we would connect the \SM to top--condensation including all loop 
corrections, we would have to use just the logarithmic RG--running 
without involving
the quadratic running of the scalar mass parameter. This would come up to the 
impossible task of reproducing the low energy phenomenology by calculating
the full dynamics of a strongly interacting gauge theory.

We conclude 
\begin{enumerate}
\item
that it is methodically wrong to include the box diagrams into the 
RG--approach and
\item
that the tree level top--condensation Lagrangian is sufficient to
make this RG--approach exact.
\end{enumerate}  
Therefore all we have to do is analyze, which relevant higher 
dimensional operators we can get from tree level heavy vector exchange.
In other words, we want to know which of these higher dimensional 
operators can be identified with operators of the effective theory.

Schematically the expansion of tree level vector exchange
 in powers of $1/M^2$ looks like
\bea
{\cal L} & = & -g \; \overline\psi \gamma_{\mu} \psi V^{\mu} - \frac{1}{4}
F_{\mu \nu} F^{\mu \nu} + \frac{M^2}{2} V_{\mu} V^{\mu} 
\nonumber \\
& = & - \frac{g^2}{2M^2} j_{\mu} j^{\mu} 
- \frac{g^2}{4M^4} J_{\mu \nu} J^{\mu \nu } + O(M^{-6}),
\eea
with
\beq
j_{\mu} = \overline \psi \gamma_{\mu} \psi, \, J_{\mu \nu} = \left(
\partial_{\mu} j_{\nu} - \partial_{\nu} j_{\mu} \right).
\label{hov}
\eeq

Now one can compare this expansion with the expansion of tree level scalar 
exchange. While the dimension four operator can be transformed into the 
corresponding scalar exchange operator by fierzing, this is not possible
for the higher dimensional operators due to their derivative couplings.
This means that tree level vector exchange does not give scalar kinetic
contributions. Boxes and higher loops, which would 
give scalar kinetic contributions,
do not contribute to the RG--analysis, as we have discussed above. Thus
these scalar kinetic contributions are just threshold effects which 
do not contribute to low energy physics. The RG--description of heavy
vector exchange connects to exactly the same set of high scale operators 
as the RG--description of minimal top condensation. The two descriptions
are therefore identical. The difference
in low energy predictions between the two models cannot be formulated 
in the framework of the renormalizable effective theory and therefore 
is not relevant.


\section{Connections to Dynamical Methods}                        
\label{dm}

In the previous sections the relevance of higher dimensional operators
was investigated with respect to the RG--approach. We found that there are
crucial differences between an underlying scalar theory and a vector 
boson exchange. These differences also have to be visible on the 
level of Schwinger--Dyson equations and it is the aim of this section to 
clarify this issue.

Even in lowest order $1/N_c$ the full theory cannot be solved. 
Thus all we can do is to investigate whether the lowest order
dynamical calculations of heavy scalar respectively vector exchange already
show the properties observed in the RG--discussion. 

We start by introducing a higher dimensional operator in the effective 
four--Fermion vertex:
  \beq   
    G = G_{\kappa=0}\left(1+\kappa\frac{p^2}{M^2}\right)
  \label{GG0}
  \eeq
If we consider scalar exchange with $G_{\kappa=0}=g^2N_c/M^2$ 
that term is contained in the expansion of the scalar propagator:
  \beq
    (-ig)\frac{i}{p^2-M^2}(-ig)=\frac{iG_{\kappa=0}}{N_c}
    \left(1+\frac{p^2}{M^2}\right) +O\left(\frac{p^4}{M^4}\right)
  \label{prex}
  \eeq
Now the crucial point is that the corresponding gap equation 
  \beq   
    1=\frac{2G_{\kappa=0}}{(4\pi)^2}\left(\Lambda^2-m^2\ln\frac{\Lambda^2}{m^2}
    \right)
  \eeq
still involves the coupling $G_{\kappa=0}$ without the additional 
contribution since there is no momentum flux into the tadpole. 
The Higgs propagator can be found by summing up the 1PI graphs
  \beq
    iD~=~\frac{-iG}{g_t^2N_c}+\frac{-iG}{g_t^2N_c}(i\Pi)\frac{-iG}{g_t^2N_c}
    +\cdots~ 
    =~ \frac{\frac{-iG}{g_t^2N_c}}{1-\Pi\frac{G}{g_t^2N_c}}~, 
  \eeq
where
  \bea
    i\Pi&=&\frac{ig_t^2N_c}{(4\pi)^2}\left(2\Lambda^2+(p^2-6m^2)\ln
    \frac{\Lambda^2}{p^2}\right) \\
    &=&i\frac{g_t^2N_c}{G_{\kappa=0}}+\frac{ig_t^2N_c}{(4\pi)^2}
    (p^2-4m^2)\ln\frac{\Lambda^2}{p^2}
  \eea
by using the gap equation.  
We find
  \bea
    iD&=&\frac{-iG}{g_t^2N_c} \left(1-\frac{G}{G_{\kappa=0}}-
    \frac{G}{(4\pi)^2}(p^2-4m^2)\ln\frac{\Lambda^2}{p^2}\right)^{-1} \\
    &=& i\left(\left(\kappa\frac{g_t^2N_c}{GM^2}+\frac{g_t^2N_c}{(4\pi)^2}
    \ln\frac{\Lambda^2}{p^2}\right)p^2-\frac{g_t^2N_c}{(4\pi)^2}4m^2
    \ln\frac{\Lambda^2}{p^2}\right)^{-1}~. 
  \eea
The zero of the Higgs self--energy and therefore the Landau pole is now 
shifted due to the different four--Fermion couplings in the propagator
and the gap equation. 
If all higher order terms are included in eq.~(\ref{prex}) and the
identification $g^2=g_t^2$ is made, one ends up with:
  \beq
    iD= i\left[\left(1+\frac{g_t^2N_c}{(4\pi)^2}
    \ln\frac{\Lambda^2}{p^2}\right)p^2-\frac{g_t^2N_c}{(4\pi)^2}4m^2
    \ln\frac{\Lambda^2}{p^2}\right]^{-1}
    \label{hipro} 
  \eeq
The zero of the Higgs self--energy and therefore the pole of the top--Yukawa
coupling $g_t$ is not at the cutoff scale since we started with a scalar theory
and a well defined finite $g_t$ at the cutoff. The Higgs self--energy 
is one
at this scale so that $g_t$ there matches the value $g$
of the underlying Lagrangian.
The pole of $g_t$ 
  \beq
    \Lambda_{\rm pol}=\Lambda e^{\frac{(4\pi)^2}{2g_t^2N_c}}
  \eeq
is beyond the cutoff.
Since we did not introduce a heavy quartic Higgs coupling in our example,
$\lambda$ must be zero at $\Lambda$. 
In the effective picture this is 
fulfilled by the Higgs mass in the propagator eq.~(\ref{hipro}).

In the case of a heavy vector exchange the situation changes drastically.
In the $1/N_c$ approximation the simple ladder appears in the Higgs
propagator as well as in the gap equation. The momentum dependence of
the four--Fermion vertex due to the higher dimensional operator in 
eq.~(\ref{GG0}) is not restricted to the s--channel. We get the same
momentum dependence also in the t--channel. The gap equation therefore 
gets the same correction as the Higgs self--energy and 
we do not get a pole shift like in case of heavy scalar exchange.
The higher dimensional
operator becomes irrelevant in the low energy limit in agreement with
the RG--argument.


\section{Conclusion}

Several questions in top--condensation which are quite difficult to handle
by dynamical methods can be answered by extending the renormalization group
(RG) approach to include higher dimensional operators. This method leads
to constituent conditions different from the simplest case if additional 
relevant higher dimensional operators are introduced. It allows a clear
and simple distinction between relevant and irrelevant operators.  While 
relevant operators in the full theory change the constituent conditions
and therefore the low energy predictions, the irrelevant operators
are responsible for distinguishing the full theory from the effective
theory near the cutoff scale. If irrelevant operators are absent, 
the model is just a reparametrization of its effective theory.
This approach gives an intuitive formulation of the argument by 
Hasenfratz et al. and Zinn-Justin
claiming that the \SM can always be reparametrized as a 
top--condensation model. It turns out that top--condensation 
by heavy scalar exchange is no theory in its 
own right but just a special case of a 
\SM reparametrization. 
The boundstate character of the Higgs has no physical significance since all
operators of top--condensation are relevant at low energies.
The specific role of the quadratic running in the
RG--approach allows also the conclusion that the masses of vectorial 
boundstates in 
top--condensation cannot be lowered below the cutoff scale. 
Finally it can be seen that the replacement of a simple 
four--Fermion coupling by heavy gauge boson exchange does not introduce new
relevant operators and therefore cannot change the
pole conditions in the RG--approach. This guarantees that top--condensation 
by heavy vector boson interaction gives well controlled predictions for the top
and the Higgs mass.

\vspace{.5cm}
{\bf Acknowledgments:} We would like to thank D. Kominis and M. Lindner
for useful discussions and M. Lindner for helpful comments
on the draft version of this paper.

This work was supported in part by the DFG under contract Li519/2-2.


\parskip=0ex plus 1ex minus 1ex


\end{document}